\documentstyle[12pt,aaspp4]{article}

\received{1995 November 27}
\accepted{}
\cpright{AAS}{}
\lefthead{GASKELL}
\righthead{ORBITAL MOTION IN A QUASAR}

\begin{document}

%
\def\deg{$^{\rm o}$}
\def\kms{\ifmmode {\rm km\ s}^{-1} \else km s$^{-1}$\fi}
\def\Msun{\ifmmode {\rm M}_{\odot} \else M$_{\odot}$\fi}
\def\Halpha{\ifmmode {\rm H}\alpha \else H$\alpha$\fi}
\def\Hbeta{\ifmmode {\rm H}\beta \else H$\beta$\fi}
\def\o{\o}
%

\title{Evidence for Binary Orbital Motion of a Quasar Broad-Line Region}

\author{C. Martin Gaskell}
\affil{Department of Physics \& Astronomy, University of Nebraska,
Lincoln, NE 68588-0111}

\begin{abstract}
Analysis of spectra of the quasar 3C 390.3 covering a 
period of over 20 yr shows that the blueshifted peak of \Hbeta\/ has
been changing its radial velocity at an almost constant rate during this time.
The radial velocity has increased by over 1500 km s$^{-1}$. The lower limit
to the period of radial velocity changes is 210 yr. Although very long
periods cannot be excluded by the radial velocity curve alone, other 
considerations suggest that the period is 
$\sim 300$ yr. 
If the radial velocity changes are
due to orbital motion, the radius of the orbit is $\sim 0.3$ pc and the total
mass of the system is $\sim 7 \times 10^9$ \Msun. In the binary black-hole model
the masses of the two holes are $\sim 2.2 \times 10^9$ \Msun\/ 
and $\sim 4.4 \times
10^9$ \Msun . A possible third peak claimed to be present in some 
1974-75 spectra is
shown to be an instrumental artifact. The narrowness of the displaced
peaks in 3C 390.3 objects requires that the broad line region cloud
motions are not governed solely by gravity. The black hole masses derived
by Koratkar \& Gaskell (1991) need to be increased significantly.
This probably lowers accretion efficiences to less than 1\% of the Eddington
limit. 
\end{abstract}

\keywords{accretion: accretion disks --- black hole physics --- galaxies: active
--- quasars: emission lines --- quasars: general --- 
quasars: individual (3C 390.3)}

\newpage

\section{INTRODUCTION}

While most quasars (by which I mean all active galactic nuclei without
regard to luminosity) have broad emission lines that are superficially 
symmetric, there is a class of quasars with broad lines showing one or two
peaks Doppler shifted from the rest frame of the host galaxy by 
substantial amounts 
(Gaskell 1983). The first example discovered was 3C 390.3 (Sandage 1966;
Lynds 1968)
and, since it is also by far the best studied case, I will refer to quasars with
displaced broad-line emission peaks as ``3C 390.3 objects''. I give their
basic properties in Gaskell (1983, 1988b). For more details than can be
discussed here, and for 
statistics of the occurence of 3C 390.3 quasars, I refer the reader to
Gaskell (1996).
The most spectacular cases tend to be radio-loud but there are also
radio-quiet cases and I argue in Gaskell (1996) that displaced broad
line peaks are very common, especially among quasars that are not seen
close to face-on.
The main 
models suggested to explain the displaced peaks include:
\begin{enumerate}
\item 
{\it Bi-conical ejection.} (Burbidge \&
Burbidge 1971; Zheng, Binette \& Sulentic 1990; Zheng, Veilleux \& 
Grandi 1991) The main problem with ejection models (Oke 1987;
Gaskell 1988b) is that 3C 390.3 objects tend {\it not} to be face-on. The gas
motion is almost certainly {\it perpendicular} to the line of sight. 
Additionally, variability studies show that the motion of BLR gas is certainly
{\it not} pure outflow (Gaskell 1988a; Koratkar \& Gaskell 1991a,b; 
Crenshaw \& Blackwell 1990; Maoz et al. 1991; Korista et al. 1995), although
generalizing from these to 3C 390.3 objects might be unwise.
\item
{\it Supermassive Binary Black Holes.} Gaskell (1983) suggested that 3C 390.3
quasars are ``spectroscopic binaries'' with each BLR peak being associated with
its own black hole. The existence of binary black holes in quasars had been
suggested by Begelman, Blandford \& Rees (1980) as an explanation of the
apparent precession of radio jets.
Since it is now clear that a large fraction of galaxies harbor massive
black holes and that mergers between galaxies are common, many such 
binaries must form (a numerical simulation of the 
merger of the nuclei of two galaxies is well illustrated in the 
videotape of 
Barnes, 1992). The precession-like wiggling of radio-jets is well established
(see Roos 1988 and Lu 1990 for extensive references). The binary black hole
model gives a natural explanation for the magnitude of the velocity shifts
of displaced peaks in 3C 390.3 quasars, for
why blue and red peaks are equally common, for why
some objects show one displaced peak while others show two, for why the broad
lines are wider in 3C 390.3 objects, and for the variability
of the peaks and continuum. These issues are all discussed at length in Gaskell
(1996).
\item
{\it Line emission from an accretion disk.} Disk geometries for the BLR have been
considered by many workers (Mathews 1982)
A disk origin for the double peaks in 3C 390.3 objects was suggested by Oke (1987), 
Perez et al. (1988), and Halpern \& Filippenko (1988).
There are many problems for disk models however (Mathews 1982; Mathews \& 
Capriotti 1985; Gaskell 1988b; Gaskell 1996). The most serious 
are that many double-peaked profiles cannot 
be fit by disk models and that the model cannot explain line profile
variability (see Gaskell 1988b, 1996). In the disk model, {\it both sides of 
the profile have to vary together in response to a change in the central
ionizing continuum} (Gaskell 1988b). They do not (Gaskell 1988b; 
Miller \& Peterson 1990). The predicted
polarization is also not seen (Antonucci, Hurt \& Agol 1995).
\end{enumerate}

\noindent
The pros and cons of these and other models (e.g., the anisotropic continuum
model of Wanders et al. 1995), and ways around these difficulties,
are discussed in more detail in Gaskell (1996). 

The supermassive binary model predicts that the wavelengths of the displaced 
peaks should change. Predicted orbital periods are of the order 
of centuries (see Gaskell 1983). Halpern \& Filippenko (1988) have argued
that the apparent absence of a change in relative radial velocity in the
3C 390.3 object Arp 102B rules out the binary black hole model, but this argument
is based on the assumption of relatively low masses for the two black holes 
in Arp 102B.
On the other hand, Veilleux \& Zheng (1991), have reported
changes in the wavelength of the blue displaced peak in 3C 390.3. After a
period when the blue peak almost completely vanished it (or another feature)
reappeared at a
longer wavelength. It is
not clear, however, whether what was seen is a long
period change or indeed whether it {\it was} a steady change in wavelength
with time at all. For another object (OQ 208 = Mrk 668), Marziani et al. (1993)
have claimed that the wavelength of a displaced peak changes with continuum
luminosity rather than time. In this 
{\it Letter} I show that the wavelength of blue peak in 3C 390.3 does 
indeed change at a
constant rate with {\it time}, as predicted by the binary black hole
model in most cases, and that this change has no connection with line or
continuum flux. 

\section{DATA AND ANALYSIS}

The biggest published set of observations of 3C 390.3 are the Lick Observatory
observations made by D. E. Osterbrock, J. S. Miller and their
collaborators. These are presented in a very convenient format by Veilleux \& 
Zheng (1991). These spectra cover the 15 yr period 1974 -- 1988. Up until 
mid-1984 the observations were made with various Robinson-Wampler Image 
Dissector Scanner (IDS) 
spectrograph systems; after mid-1984 observations
were made with a CCD spectrograph. The majority of the Lick spectra are of the
\Hbeta\/ region.

Veilleux \& Zheng (1991) only give eye estimates of the wavelength of the
blue peak. I therefore measured all of the Lick spectra that showed a convincing 
blue displaced peak. 
A quasi-continuum was set under the displaced peak to allow for blending
with the central (undisplaced) emission line.
Rather than using the apparent ``peak'' of the displaced line, which is 
noise-sensitive, the 
Pogson method (Hoffmeister, Richter \& Wenzel 1985) was used to determine 
the position of each peak.
Based on a comparison of spectra taken close together, the 
rms error for a single measurement was $\pm2.0$ \AA\/
(versus $\pm 4.3$ \AA\/ for the Veilleux \& Zheng measurements). My measurements
of the peak positions were systematically blueshifted by 4.0 \AA\/ relative to 
the Veilleux \& Zheng eye estimates. During the ``low state'' of 3C 390.3 
(circa 1980),
when the blue peak was too weak to measure, I adopted the wavelengths of Veilleux
\& Zheng, but blueshifted them by 4 \AA\/ to put them on the same system as 
my measurements. 

The 1965 photographic 
spectrum of Sandage (1966) has a small defect right where the center of the
peak is predicted to be. It cannot be used for a precise measurement, but
is useful for confirming that the blue peak existed 30 years ago. Lynds (1968)
presents a number of photographic image-tube spectra taken in 1967. The 
positions of the blue peaks in these were measured in enlargements. The
uncertainty of the average wavelength was calculated from the residuals of
all lines from the wavelength calibration curve. Burbidge \& Burbidge (1971)
give measurements of several 1970 Lick Observatory photographic image
tube spectra. I adopted the mean \Hbeta\/ blue peak wavelength from their table 1.
The error bar was calculated as for the Lynds (1968) spectrum.

3C 390.3 also shows a {\em redshifted} displaced peak 
which can be quite prominent
at times (e.g., when the continuum and blue peak were faint in June/July 1980).
It would obviously be interesting to also get relative radial velocities for 
the redshifted peak. There are two general problems however. First, this peak,
unfortunately, is twice as close to the unshifted
peak as the blue one is, so blending is more of a problem and the position depends
on how the central peak is subtracted off. Second, there is also blending
with the strong [O III] $\lambda$4959 line. Nevertheless, Veilleux \& Zheng (1991)
do tabulate eye estimates of wavelengths of the red peak. 
They indicate that these are highly
uncertain. Because of the difficulty in measuring this peak, no attempt has
been made here to measure the peak wavelengths and no radial velocity curve has
been given.

Veilleux \& Zheng claim there is
a {\em third} peak, redshifted by + 4600 \kms, which is only visible in the Lick
spectra of 
1974-1975. If this were real it would pose a serious problem for many models. 
However, weak features in IDS spectra need to be treated with caution for a 
number of reasons. First, the IDSs
were not photon-counting detectors. Each photon
produced a different number of counts (see Robinson \& Gaskell 1978). Second,
the illumination of the detector by the quartz lamp continuum
source used to divide out small-scale response variations was frequently
different than the 
the illumination by the target object. This could produce repeatable
spurious features.
Different observers used different slits,
deckers and apertures, even on the same IDS. Finally, the IDS was heavily
oversampled, so artificial bumps in spectra {\it look} smooth.
Examination of the 3C 390.3 spectra shows that the ``third'' peak
is almost certainly a result of the IDS
instrumental problems. Rather than being a true emission
feature, the + 4600 \kms\/ feature is really the
result of a sharp {\em dip} in the spectrum to the blue. This blue dip is most
prominent in two spectra taken by the Miller group on June 4 and June 5 1975. The
instrumental nature of this is demonstrated by it
vanishing in a spectrum taken by the Osterbrock group with 
a different instrumental
setup {\em on the very next night}. 

\section{RESULTS}

The radial velocity curve for the blue peak is shown 
in Figure 1. Points are annual means. Error bars are $\sigma / \surd n$. 
The open
circles are median values (from Veilleux \& Zheng 1991) when the blue peak was
essentially absent from the spectra. The latter have not been used in any
further analysis. 

The most striking result is the almost linear change in the radial velocity
of the blue peak over more than two decades. The brightness of 3C 390.3 has 
varied a lot during this period 
and the character of the variability has
changed as well. Shen, Usher \& Barrett (1972) report a
change of 1.15 magnitudes in only 3 days, suggesting 3C 390.3 was in the class
of optically violent variables (OVVs) at that time. The high polarization also
supports an OVV classification (see Angel \& Stockman 1980).
The Herstmonceux monitoring of 
Lloyd (1984)
shows a similarly rapid, though not quite as large, change in 1969. The
Herstmonceux light curve shows that 3C 390.3 was much brighter and more
active around the
time of the Lynds (1968) and Burbidge \& Burbidge (1971) spectra than 
during most of the Lick spectra.
An important implication of Figure 1
is that {\em the change in radial velocity is not a function of the
continuum luminosity.} Marziani et al. (1993) claimed that the wavelength
of one of the displaced
peaks in OQ 208 correlated with luminosity. This is clearly not
the case with the blue peak in 3C 390.3 since the wavelength is increasing
monotonically while the luminosity of both the continuum and the blue peak
goes through a minimum around 1980. Since the offset peak in OQ 208 is
blended with the central unshifted peak, I believe that the 
correlation Marziani et al. found is caused by changing blending 
with the central peak.
The correlation they find is certainly in the right sense and of the
right magnitude for this to be the cause. In 3C 390.3 it is obvious that
the steady change in wavelength has persisted from when
3C 390.3 was bright and ``active'', through the ``low state'', around 1980 when
the peak almost completely vanished, and on to the revival in the later 1980's.

We can draw at least two conclusions from this:
\begin{enumerate}
\item
The gas producing the blue peak is long-lived. It is not some transient
feature. The peak is wide enough that over 20 yr it would have 
dissipated if the gas were not bound to something or if there were no source
of gas.
\item
The wavelength of the peak has nothing to do with the luminosity of the
central engine. Since, as is normal for quasars, the total \Hbeta\/ 
and \Halpha\/ fluxes
track the observed continuum
luminosity (see Figure 1 of Oke 1987 and Figures 1a and 1b of Veilleux \&
Zheng 1991), it is not possible to claim that the variability of the
ionizing continuum emission
is highly anisotropic and that the continuum variation observed on earth is radically
different from that seen by the blue peak.
\end{enumerate}

We can learn more from Figure 1 if we adopt the natural hypothesis that 
the change in radial velocity is the
result of orbital motion of the gas producing the blue peak. The first step
is to determine the orbital period. It is probably fairly safe to assume that
the orbit is circular: dynamic friction will probably have accomplished this.
Unfortunately, the period is not well constrained. A straight-line 
(infinite period) already has a $\chi ^2$ per degree of freedom close to
unity. It is possible, however, to set a lower limit on the period. A period
of 210 yr (the most extreme curve in Figure 1) is excluded at the 90\%
confidence level. A period of 300 yr (the middle curve in Figure 1) is
completely acceptable ($\chi ^2$ per d.o.f. = 0.8). 
A period of 210 yr gives a maximum
observed velocity of 4700 \kms ; a period of 300 yr gives 5340 \kms .
For long orbital periods the maximum velocity scales approximately linearly 
with the orbital
period. Since the observed relative velocity of the blue peak is already one
of the largest known, orbital periods at the lower end of the acceptable
range are favored. I will therefore
adopt a period of 300 yr for further calculation. The true period will
probably be within 50\% of this.

It is obvious from the curves in Figure 1 that it will be possible within
a few years to constrain the orbital period quite tightly (possibly through
the current {\it International AGN Watch} monitoring program). Measurement
of possible additional archival 1960's spectra would also be helpful. The 
predicted date of zero relative radial velocity is insensitive to the period.
If the period is 210 yr, zero relative radial velocity is predicted to 
occur in the year 2012; if the period is
300 yr, it will be in 2018 (the limiting case of an infinite period gives
2029). In the second decade of the 21st century 3C 390.3 should no longer
be a 3C 390.3 object!

Adopting a 300 yr period, the maximum velocity is 4700 \kms . For a circular
orbit this is $v_{max} \sin i$, where $i$ is the inclination of the 3C 390.3 jets
to our line of sight. This inclination can be estimated from the 
synchrotron self-Compton model and relativistic beaming considerations 
(Ghisellini et al. 1993). The ratio of optical flux to radio core flux supports
such estimates (Wills \& Brotherton 1995). For 3C 390.3, Ghisellini
et al. get $i \sim 29$\deg. This gives $v_{max} \sim 9400$ \kms . The radius
of the orbit is therefore almost exactly one lt-yr (0.3 pc) and the total
mass is $6.6\times 10^9$\/ \Msun . The ratio of the
relative radial velocity shifts of the blue and red peaks is 2:1 if we assume
that the narrow lines indicate the combined center of mass, so the masses
of the black holes are $4.4\times 10^9~\Msun$ and $2.2\times 10^9~\Msun$.

If the binary black hole model is correct, the redshifted peak should be
showing the opposite motion to the blueshifted peak. As mentioned in the
previous section there are major difficulties in determining the relative
radial velocity of the red peak, but the changes in the wavelength of 
the red peak do indeed seem to be consistent with the predictions of the
binary model. Over the last 10 years of the Lick data, the magnitude of the
relative radial velocity of the red peak has decreased by many hundreds of \kms.

\section{DISCUSSION}

Although the black hole masses found are 
in agreement with the masses claimed for black holes in nearby galaxies 
(e.g., the {\it HST} determination of $2.4 \times 10^9$\Msun\/ in M 87; 
Ford et al., 1994), 
the black holes masses found are somewhat higher than predicted
by the Koratkar \& Gaskell (1991b) mass-luminosity relationship. A possible 
reason
for this is easy to understand. Koratkar \& Gaskell obtained their quasar mass
estimates from velocity dispersions and size estimates from their
``reverberation mapping''. Gaskell (1988a) had also shown that
the motion of BLR gas is not predominantly radial, and Koratkar \&
Gaskell (1991a) found this to be generally true of the quasars they
considered. They therefore assumed that the motions were gravitationally
dominated and used the virial theorem to find the masses. Penston (1987)
pointed out that gravitationally dominated BLR motion was incompatible 
with the binary
black hole model because one needs velocity dispersions {\it lower}
than the orbital velocity for gas bound to each black hole (for 
discussion of this point see footnote 3 of Cheng, Halpern \& Filippenko 1989).
There is a natural solution to this apparent dilemma: 
unlike black holes,
BLR clouds can easily be influenced by non-gravitational forces.
Both gravity and radiation pressure are inverse square law forces. Clouds
can orbit under the combined force. The mean velocity at a given radius
from a black hole will be less than the orbital velocity of particles 
subject only to gravity. If one assumes that the BLR clouds are subject
only to gravity a serious underestimate of the black hole mass results.
There is every reason to believe that radiation pressure and wind pressures
are {\it very} significant and that in some situations they can even exceed
gravity, since at least some gas is being seen to be
expelled at high velocities from quasars (the broad absorption line gas).
A more complicated model invoking both radiation pressure and gravity
has been presented by Mathews (1993). He estimates that Koratkar \& Gaskell
(1991b) have underestimated black hole masses by a factor of about 10--20.

If this reconciliation of reverberation mapping results with the binary
black hole model is correct, then the Eddington efficiences found by
Koratkar \& Gaskell (1991b) have also been overestimated. Quasar black
hole accretion
efficiences fall from a few percent to a few tenths of a percent.

The size of the orbit is only a
few times the size of the BLR inferred from reverberation mapping.
Obviously cloud motions in the presence of two black holes are complicated
and detailed modelling work is needed. 

The spin axis of a rotating orbiting black hole can be subject to geodetic
precession. Given the masses and orbital sizes, the precession period,
$P_{prec}$,
can be derived (see equation 8 of Begelman, Blandford \& Rees 1980). For
the parameters just inferred from the 3C 390.3 radial velocity curve, $P_{prec}
\sim 4 \times 10^5$ yr.
This is very typical of precession
periods deduced from wiggles in quasar jets (see Roos 1988 and Lu 1990). The jet of 3C 390.3
is straight however, at least on kpc scales. This implies that, in this
case, the spin and orbital angular 
momentum vectors are closely parallel.

Although the simple disk model of 3C 390.3
objects has been strongly ruled out for a long time (see introduction), 
there have been modifications to the model.
Zheng, Veilleux \& Grandi (1991) introduced ``hot spots'' on the
disk (there is precedence for this from cataclysmic variable star
work). The radial velocity change in Figure 1 is compatible with
a hot spot orbiting one black hole, but such a model does not explain
why the blue peak is correlated with the continuum changes while the
red one is not, or indeed why the profile should change at all.
Eracleous et al. (1995) introduced elliptical accretion disks to
explain some profiles not fit by simple accretion disk models.
In these models the observed radial velocity change
would be due to relativistic precession and tidal effects of a
binary companion black hole. The period is about right, but, as
Eracleous et al. concede, this model does not explain changes in the
intensity ratios of the red and blue peaks. 
Chakrabarti \& Wiita (1994) propose that the humps are
due to spiral shocks in accretion discs. Unfortunately, they
predict a period of 21 yr for 3C 390.3, which is too short by at
least a factor of 10, and, once again, the model fails to explain
the line profile and continuum variability.

\section{CONCLUSIONS}

The observations analyzed here present strong support for the binary
black hole model. One of the major predictions of the model, a
change in radial velocity over time, has been
accurately confirmed. The initial failure to detect such a change in
Arp 102B is not a problem if Arp 102B has black hole masses as large
as in 3C 390.3 or in a typical remnant. It is {\it possible} that the 
radial velocity
changes seen in Figure 1 are produced by a ``hot spot'' or irregularity
in a disk structure, but a number of factors (the shape of the line
profile and the nature of the line variability in 3C 390.3
itself, and 3C 390.3 objects in general) argue strongly against this model.
The production of supermassive black hole binaries through galactic
mergers is something we know {\it must} be going on. 3C 390.3 objects seem
to be a likely consequence. Clearly it will be interesting to follow
other 3C 390.3 objects to see whether they too show radial velocity
variations similar to those found in 3C 390.3 itself. Sergeev et al. (1994)
have already claimed a velocity change of 2400 \kms\/ in $\sim 800$ d
in NGC 5548.

I am grateful to Ski Antonucci, Mike Eracleous, Don Osterbrock, Sylvain
Veilleux, Ignaz Wanders, and Wei Zheng for discussion of this work.

%

%
%
%
\newpage
\begin{figure}
\plotfiddle{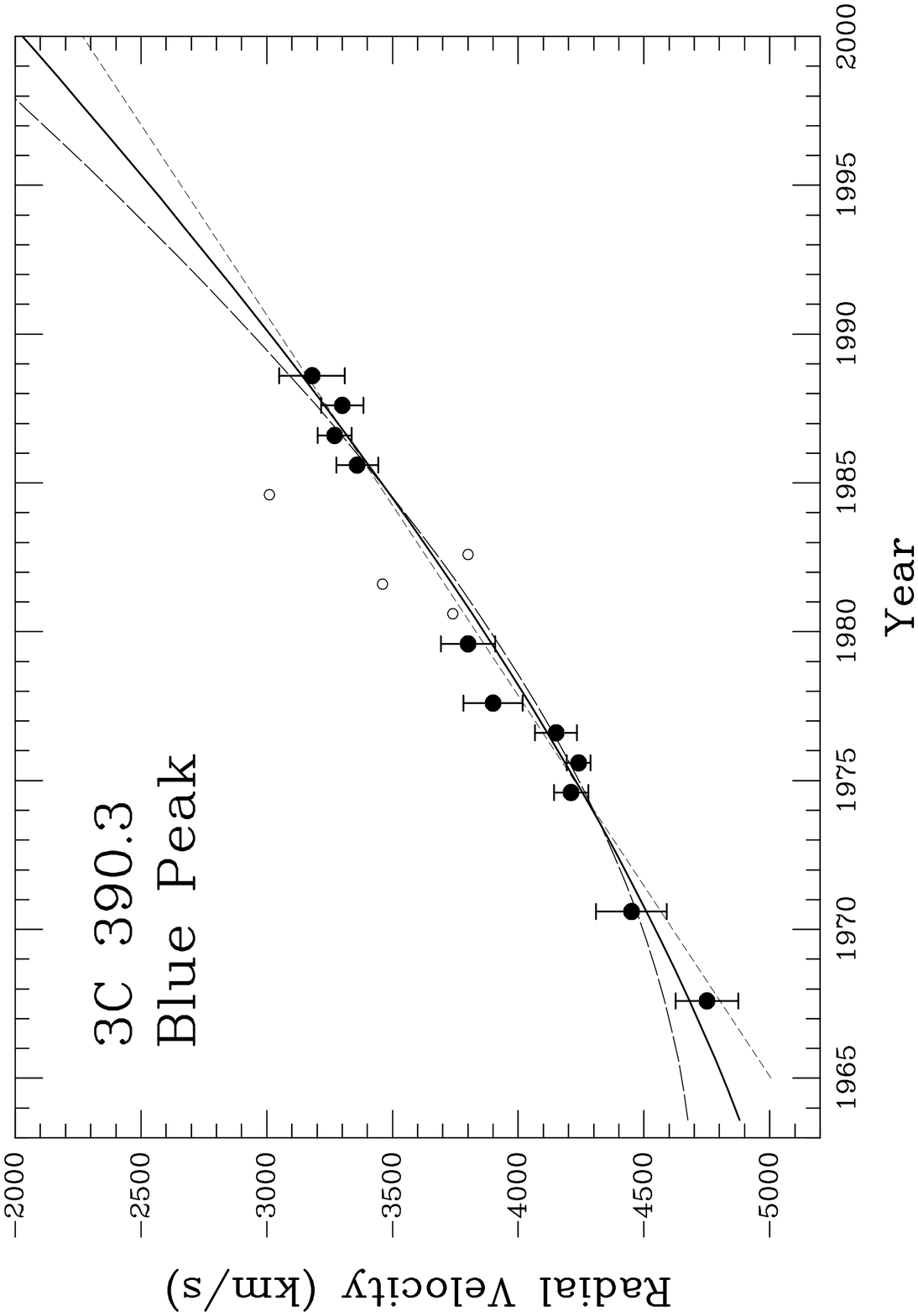}{4.5in}{270}{64}{64}{-252}{432}
\caption{
\noindent
Radial-velocity curve for the blue displaced broad peak of H$\beta$
in 3C 390.3. Error bars are calculated as described in the text. The
open circles are median values from the eye estimates of Veilleux \&
Zheng (1991) when the blue peak was essentially absent from the spectra.
The long-dashed curve
is for a 210 yr period (excluded at the 90\% confidence level), 
the solid curve is for a 300 yr period and the straight line is 
the limiting case of an infinite period.
}
\end{figure}

\newpage

\begin{references}

\reference{} Angel, J. R. P. \& Stockman, H. S. 1980, ARA\&A, 8, 321
 
\reference{} Antonucci, R. R. J., Hurt, T. \& Agol, E. 1995, \apj\/ Letters, in 
press

\reference{} Barnes, J. E. 1992, \apj, 393, videotape segment 2

\reference{} Begelman, M. C., Blandford, R. D., \& Rees, M. J. 1980,
Nature, 287, 307

\reference{} Burbidge, E. M. \& Burbidge, G. R. 1972, \apj, 163, L21

\reference{} Chakrabarti, S. K. \& Wiita, P. J. 1994, \apj, 434, 518

\reference{} Cheng, K., Halpern, J. P. \& Filippenko, A. V. 1989 \apj, 339, 742

\reference{} Crenshaw, D. M. \& Blackwell, J. H., Jr. 1990, \apj, 358, L37

\reference{} Eracleus, M., Livio, M., Halpern, J. P., \& Storchi-Bergmann, T.
1995, \apj, 438, 610

\reference{} Ford, H. C. et al. 1994, \apj, 435, L27

\reference{} Gaskell, C. M. 1983 in Quasars and Gravitational Lenses,
24th Liege Astrophysical Colloquium, 471

\reference{} Gaskell, C. M. 1988a, \apj, 325, 114

\reference{} Gaskell, C. M. 1988b, in Active Galactic Nuclei, ed. H. R.
Miller \& P. J. Wiita (Berlin: Springer), 61

\reference{} Gaskell, C. M. 1996, in Jets from Stars and Active Galactic Nuclei, ed. 
W. Kundt, (Berlin: Springer), 165 (astro-ph/9605175)

\reference{} Ghisellini, G., Padovani, P., Celotti, A., \& Maraschi, L.
1993, \apj, 407, 65

\reference{} Halpern, J. P. \& Filippenko, A. V. 1988, Nature, 331, 46

\reference{} Hoffmeister, C., Richter, G., \& Wenzel, W. 1985, Variable Stars
(Berlin: Springer),  280

\reference{} Koratkar, A. P. \& Gaskell, C. M. 1991a, \apjs, 75, 719

\reference{} Koratkar, A. P. \& Gaskell, C. M. 1991b, \apj, 370, L61

\reference{} Korista, K. T., et al.\ 1995, \apjs, 97, 285

\reference{} Lloyd, C. 1984, \mnras, 209, 697

\reference{} Lu, J. J. 1990, \aap, 229, 424

\reference{} Lynds, C. R. 1968, \aj, 73, 888

\reference{} Marziani, P., Sulentic, J. W., Calvani, M., Perez, E., Moles, 
M. \& Penston, M. V. 1993, \apj, 410, 56

\reference{} Maoz, D., Netzer, H., Mazeh, T., Beck, S., Almoznino, E.,
Leibowitz, E., Brosch, N., Mendelson, H. \& Laor, A. 1991, \apj, 367, 493

\reference{} Marziani, P., Sulentic, J. W., Calvani, M., Perez, E., \&
Penston, M. V. 1993, \apj, 410, 56

\reference{} Mathews, W. G. 1982, \apj, 258, 425

\reference{} Mathews, W. G. 1993, \apj, 412, L17

\reference{} Mathews, W. G. \& Capriotti, E. R. 1995, in Astrophysics of
Active Galaxies and Quasi-Stellar Objects, ed. J. S. Miller (Mill Valley:
University Science Books), 185

\reference{} Miller, J. S. \& Peterson, B. M. 1990, \apj, 361, 98

\reference{} Oke, J. B. 1987, in Superluminal Radio Sources, ed. J. A. 
Zensus \& T. J. Pearson (Cambridge: Cambridge Univ. Press), 267

\reference{} Penston, M. V. 1987, private communication

\reference{} Perez, E., Penston, M.V., Tadhunter, C., Mediavilla, E., \& Moles,
M. 1988, \mnras, 230, 353

\reference{} Robinson, L. B. \& Gaskell, C. M. 1978, in 7th Symposium on
Photoelectronic Imaging Devices (London: Imperial College), 103

\reference{} Roos, N. 1988, \apj, 334, 95

\reference{} Sandage, A. R. 1966, \apj, 334. 95

\reference{} Sergeev, S. G., Malkov, Yu. F., Chuvaev, K. K., \& Pronik, 
V. I. 1994, in Reverberation Mapping of the Broad Line Region in Active
Galactic Nuclei, ed. P. M. Gondhalekar, K. Horne, \& B. M. Peterson
(San Francisco: Astronomical Society of the Pacific), 199

\reference{} Shen, B. S. P., Usher, P. D. \& Barrett, J. W. 1972, \apj, 171, 457

\reference{} Veilleux, S., \& Zheng, W.\ 1991, \apj, 377, 89

\reference{} Wanders, I., et al. 1995, ApJ 453, L87

\reference{} Wills, B. J. \& Brotherton, M. S. 1995, \apj, 448, L81

\reference{} Zheng, W., Binette, L., \& Sulentic, J.W. 1991, \apj, 365, 115

\reference{} Zheng, W., Veilleux, S., \& Grandi, S.A. 1991, \apj, 381, 418

\end{references}
\end{document}